\documentclass[10pt, conference]{IEEEtran}
\IEEEoverridecommandlockouts
\usepackage{float}
\usepackage{subcaption}
\usepackage{url} 
\usepackage{makecell}
\usepackage[most]{tcolorbox}
\usepackage{placeins}
\usepackage{pifont}
\newcommand{\checkmark}{\ding{51}}
\newcommand{\xmark}{\ding{55}}
\usepackage{cite}
\usepackage{amsmath,amssymb,amsfonts}
\usepackage{algorithmic}
\usepackage{graphicx}
\usepackage{dblfloatfix}
\usepackage{textcomp}
\usepackage{xcolor}
\def\BibTeX{{\rm B\kern-.05em{\sc i\kern-.025em b}\kern-.08em
    T\kern-.1667em\lower.7ex\hbox{E}\kern-.125emX}}
\begin{document}

\title{A Multi-Level Integrity Evaluation Framework for Quantum Circuits under Controlled Anomaly Injection}

\author{
\IEEEauthorblockN{
Ejaz Ahmed\IEEEauthorrefmark{1},
Boshuai Ye\IEEEauthorrefmark{1},
Syed Hamza Shah\IEEEauthorrefmark{1},
Muhammad Azeem Akbar\IEEEauthorrefmark{2},
Arif Ali Khan\IEEEauthorrefmark{1}
}
\IEEEauthorblockA{
\IEEEauthorrefmark{1}\textit{ITEE, University of Oulu}, Oulu, Finland\\
\{Ejaz.Ahmed, boshuai.ye, Hamza.Shah, Arif.khan\}@oulu.fi
}
\IEEEauthorblockA{
\IEEEauthorrefmark{2}\textit{Software Engineering Department, Lappeenranta-Lahti University of Technology}, Lappeenranta, Finland\\
Azeem.Akbar@lut.fi
}
}

\maketitle

\begin{abstract}
To ensure the integrity of quantum circuits is a significant challenge in the Noisy Intermediate-Scale Quantum (NISQ) era, where circuits are subject to compilation transformations, hardware constraints, and potential adversarial modifications. Existing validation approaches typically rely on either structural analysis or behavioral evaluation, leading to incomplete assessment of circuit correctness.

In this work, we investigate the relationship between structural, interaction-level, and behavioral perspectives of circuit integrity, demonstrating that single aspect of integrity is insufficient to guarantee circuit integrity; structural similarity alone does not ensure behavioral equivalence. To solve this problem, we use a three-layer metric framework that combines the Structural Integrity Score (SIS), the Operational Integrity Score (OIS), and the Interaction Graph Semantic-Logical (IGS). SIS captures global structural properties,OIS quantifies behavioral divergence using Jensen-Shannon distance, and IGS models interaction patterns and dependencies in a pre-execution setting.

Through controlled anomaly injection on benchmark quantum circuits, we demonstrate that each metric captures a different aspect of circuit deviation. In particular, structural blind-spot cases (SIS $\geq 0.95$) reveal a clear limitation of structural analysis, where OIS detects anomalies in 93.85\% of instances, while IGS detects 72.58\%. These results highlight that the metrics provide complementary insights and that a single metric is insufficient for reliable circuit validation.

\end{abstract}


\section{Introduction}

Quantum computing has progressed rapidly in the Noisy Intermediate-Scale Quantum (NISQ) era, where practical applications are executed on imperfect hardware and complex software toolchains \cite{boixo2018characterizing,arute2019quantum,preskill2018nisq}. In such environments, ensuring that a quantum circuit behaves as intended is a real challenge. Circuit instances may deviate from their original specification due to compilation transformations, optimization passes, hardware constraints, and unintended or adversarial modifications \cite{zulehner2018mapping,maslov2008quantum}. Since quantum hardware is not widely accessible, workflows depend on cloud-based execution and hybrid classical quantum pipelines, increasing the need for circuit-level validation. Existing validation approaches typically fall into two categories: structural methods, which analyze properties such as gate composition, circuit depth, and interaction topology to provide efficient pre-execution checks \cite{maslov2008quantum,amy2013meet}, and behavioral methods, which assess output distributions or state fidelity through execution or simulation to provide a direct measure of functional correctness \cite{nielsen2010quantum}. Although both perspectives are valuable and often treated independently, our results show that this leads to an incomplete assessment of circuit integrity.

In this work, we identified that structural similarity does not guarantee behavioral equivalence. Transformations such as gate reordering and gate substitution can preserve global structural properties while significantly altering circuit behavior, revealing a structural-behavioral gap. This observation highlights the need for validation approaches that account for multiple levels of circuit representation rather than relying on a single perspective. To address this, we adopt a multi-layer integrity framework that explicitly distinguishes between structural, interaction-level, and behavioral views. SIS captures global structural properties such as gate count, depth, and topology without requiring execution, OIS measures behavioral divergence using Jensen-Shannon \cite{lin1991divergence} distance between output distributions, and IGS provides an intermediate pre-execution perspective by modeling interaction patterns, dependencies, and gate-level relationships through graph-based representations. The definition and selection of these metrics are grounded in established approaches to circuit analysis, and their theoretical basis and design rationale are summarized in Table~\ref{tab:metric_foundations}.

However, existing approaches rarely examine whether structural, interaction-level, and behavioral metrics agree or diverge under the same controlled perturbations. Through controlled anomaly injection across benchmark circuits \cite{li2023qasmbench}, we investigate how these metrics capture complementary aspects of circuit integrity. SIS reliably identifies coarse structural changes, OIS consistently reflects behavioral divergence, and IGS captures interaction-level inconsistencies that are not reflected in global structural descriptors. Correlation analysis further shows that interaction-level similarity does not fully determine behavioral equivalence, reinforcing the independence of these perspectives. We make the following contributions: (i) a benchmark-driven empirical analysis of the structural-behavioral gap in quantum circuit validation, and (ii) a multi-layer integrity framework combining SIS, OIS, and IGS, together with a systematic evaluation of their complementary roles and limitations.

\section{Background and Related Work}
\label{sec:background}
\subsection{Background}

Quantum circuits executed in the NISQ era are subject to multiple sources of variation, including hardware noise, calibration drift, compilation and transpilation transformations, and execution in hybrid classical--quantum workflows \cite{boixo2018characterizing,arute2019quantum}. As a result, the same logical circuit may appear in different structural forms throughout its lifecycle, depending on optimization passes, hardware constraints, and execution environments. Backend-specific mapping and hardware variability can influence how logical circuits are transformed before execution, motivating validation methods that account for more than a single structural descriptor \cite{murali2019noise,zulehner2018mapping}. This makes circuit-level validation challenging, as structural changes introduced during processing do not necessarily imply changes in functional behavior, and conversely, behaviorally significant deviations may arise even when structural properties appear largely preserved \cite{burgholzer2021advanced}.

To support evaluation across diverse workloads, benchmark suites such as QASMBench provide standardized circuit collections together with structural and execution-related descriptors, including gate count, circuit depth, and interaction patterns \cite{li2023qasmbench,tomesh2022supermarq,tripathi2025benchmarking}. These descriptors are widely used to estimate circuit complexity, resource requirements, and hardware suitability. In parallel, behavioral evaluation methods assess output distributions or state fidelity through simulation or execution, providing a direct measure of functional correctness \cite{nielsen2010quantum}. These structural and behavioral approaches make up most of the existing methods used to assess quantum circuits. 

\subsection{Related Work}

Prior work has examined quantum circuit correctness and variability from structural, behavioral, equivalence-checking, and graph-based perspectives. Quantum circuit equivalence checking has shown that circuits generated through compilation, decomposition, and mapping may differ significantly in structure while still representing the same logical functionality \cite{burgholzer2021advanced}. This highlights the difficulty of ensuring consistency across heterogeneous circuit representations, particularly in complex design flows.

Structural analysis techniques are widely used to characterize circuit properties such as gate count, depth, and interaction topology, and to estimate complexity and hardware suitability \cite{zulehner2018mapping,li2023qasmbench}. Behavioral approaches instead evaluate circuits based on output distributions or state fidelity obtained through simulation or execution, providing a direct measure of functional correctness \cite{lin1991divergence,liu2018quantum}. In addition, fault-based studies have explored the detection of missing or incorrect quantum operations \cite{bera2015diagnosis}, while compilation and optimization research has investigated how circuit transformations such as simplification, decomposition, and reordering affect circuit structure and performance \cite{maslov2008quantum,amy2013meet}. Graph-based representations have also been proposed to model dependency structure and interaction patterns in quantum circuits \cite{itoko2020optimization,hagberg2008networkx}.

Although these approaches provide valuable insights, they are typically applied separately and focus on a single aspect of circuit correctness. Structural methods are efficient but may overlook semantic or behavioral deviations, behavioral methods provide accurate validation but require costly execution or simulation, and graph-based approaches are rarely integrated with explicit behavioral analysis. Similarly, fault-based methods are often limited to predefined fault models and are not designed for broader integrity validation under diverse perturbations. This creates a gap: circuit integrity in practical workflows must be assessed across structural, interaction-level, and behavioral representations simultaneously.

\begin{table}[t]
\caption{Summary of existing circuit validation approaches and their limitations.}
\label{tab:gap_analysis}
\centering
\begin{tabular}{p{2.3cm} p{2.7cm} p{2.7cm}}
\hline
\textbf{Approach} & \textbf{What is Captured} & \textbf{Limitation} \\
\hline
Structural Analysis 
& Circuit complexity and structure (gate count, depth, topology) \cite{zulehner2018mapping,li2023qasmbench} 
& Does not capture behavioral deviation or semantic differences caused by transformations \\

Behavioral Evaluation 
& Output distributions and functional correctness \cite{lin1991divergence,liu2018quantum} 
& Requires execution/simulation; does not explain structural or interaction-level causes \\

Fault-Based Diagnosis 
& Detection of specific circuit faults (e.g., missing or incorrect gates) \cite{bera2015diagnosis} 
& Limited to predefined fault models; not suitable for complex or compound anomalies \\

Compilation / Transformation Studies 
& Circuit optimization and transformation effects \cite{maslov2008quantum,amy2013meet} 
& Focus on optimization rather than anomaly detection or integrity validation \\

Graph-Based Representations 
& Dependency structure and interaction modeling \cite{itoko2020optimization,hagberg2008networkx} 
& Typically not integrated with behavioral validation or integrity assessment \\
\hline
\end{tabular}
\end{table}

\textbf{Motivation scenario and problem.} Consider a team developing quantum circuits for deployment on cloud-based platforms. As part of their workflow, circuits are compiled, optimized, and executed across different environments. Before deployment, the team needs to verify the integrity of these circuits to ensure that no unintended modifications have been introduced during processing.

However, existing approaches do not commonly integrate structural, interaction-level, and behavioral perspectives within a single circuit-level evaluation workflow. Some approaches focus only on structural properties, others focus on behavioral validation, while some are primarily concerned with optimization rather than integrity. As summarized in Table~\ref{tab:gap_analysis}, existing approaches capture complementary aspects of circuit behavior, including structural properties, output-level correctness, and interaction-level representations. However, these aspects are not integrated within a single framework.

In practical workflows, particularly those involving automated compilation, optimization passes, or shared execution environments, circuits may undergo transformations that are not fully reflected within a single perspective. This lack of a unified approach makes it difficult for teams to ensure circuit integrity both structurally and operationally, raising an important question: \emph{has a given circuit instance deviated from its intended structure or behavior in a way that affects its integrity?}

To overcome these limitations, a multi-metric validation strategy is required that combines structural, behavioral, and interaction-level perspectives. This motivates the central research question of this paper: \emph{how do structural, interaction-level, and behavioral metrics differ in their ability to detect controlled quantum circuit anomalies?} Addressing this question requires circuit-level integrity metrics that operate across multiple representations, combining structural descriptors, interaction-level information, and behavioral observations to provide a more comprehensive and interpretable assessment of circuit deviations.

\section{Metric Development}
\label{sec:metrics}

This section defines the three metrics used in the proposed framework: the SIS, the OIS, and the IGS. Each metric compares a test circuit $C$ against a reference circuit $C_{\text{ref}}$ and captures a different aspect of circuit deviation, namely structural, behavioral, and interaction-level integrity. Table~\ref{tab:metric_foundations} summarizes the theoretical basis and motivation behind each metric, clarifying how they are defined and adopted in this work.

\subsection{Structural Integrity Score (SIS)}

The Structural Integrity Score measures deviation in global circuit structure using normalized differences in key structural properties. These properties include gate count, circuit depth, two-qubit gate usage, and interaction topology, which are widely used in quantum circuit analysis and compilation studies \cite{zulehner2018mapping,li2023qasmbench}.

The structural deviation is defined as:

\begin{equation}
\Delta_{\text{struct}}(C, C_{\text{ref}}) =
w_g \delta_{\text{gate}} +
w_d \delta_{\text{depth}} +
w_c \delta_{\text{2q}} +
w_t \delta_{\text{topo}}
\end{equation}

where each component $\delta$ represents a normalized relative difference with respect to the reference circuit, and the weights satisfy $w_g + w_d + w_c + w_t = 1$. In this work, we assign equal weights 
($w_g = w_d = w_c = w_t = 0.25$), 
because there is no prior evidence indicating that one structural component should be prioritized over the others for general anomaly detection across diverse circuit types.

The Structural Integrity Score is then defined as:

\begin{equation}
\mathrm{SIS}(C, C_{\text{ref}}) = 1 - \Delta_{\text{struct}}(C, C_{\text{ref}})
\end{equation}

The four structural components capture complementary aspects of circuit construction: gate count reflects circuit size and detects insertions or deletions, circuit depth captures sequential complexity, two-qubit gate usage represents changes in entangling operations, and topology encodes dependency relationships among quantum operations, where edges capture ordering constraints induced by shared-qubit usage and gate commutation rules \cite{itoko2020optimization}. Collectively, these features provide a compact pre-execution measure of structural similarity. However, as SIS aggregates global properties, it is less sensitive to fine-grained semantic or interaction-level changes, particularly under structure-preserving transformations introduced during circuit optimization and routing \cite{cowtan2019qubit}.




\subsection{Operational Integrity Score (OIS)}

The Operational Integrity Score measures behavioral deviation between two circuits
using their output probability distributions. Let $p$ and $q$ denote the output
distributions obtained from executing or simulating $C_{\mathrm{ref}}$ and $C$,
respectively. To quantify behavioral difference, we use the Jensen--Shannon
distance because it is symmetric, bounded, and suitable for comparing finite
measurement-output distributions \cite{lin1991divergence,kullback1951information}.

The Kullback--Leibler (KL) divergence between two discrete distributions $p$ and
$q$ is defined as:
\begin{equation}
D_{\mathrm{KL}}(p \parallel q) =
\sum_i p(i)\log_2\frac{p(i)}{q(i)} .
\end{equation}

Let $m = \frac{1}{2}(p+q)$. The Jensen--Shannon divergence is:
\begin{equation}
D_{\mathrm{JS}}(p,q) =
\frac{1}{2}D_{\mathrm{KL}}(p \parallel m)
+
\frac{1}{2}D_{\mathrm{KL}}(q \parallel m).
\end{equation}

The Jensen--Shannon distance is then:
\begin{equation}
JSD(p,q) = \sqrt{D_{\mathrm{JS}}(p,q)} .
\end{equation}

In this work, logarithms are computed with base 2, making $JSD(p,q)$ bounded in
$[0,1]$. The Operational Integrity Score is defined as:
\begin{equation}
OIS(C,C_{\mathrm{ref}}) = 1 - JSD(p,q).
\end{equation}

This formulation maps output-distribution distance into an interpretable
similarity score, where values closer to 1 indicate stronger behavioral agreement
and values closer to 0 indicate larger behavioral deviation.

OIS directly reflects observable functional differences and serves as a behavioral reference metric. It captures deviations arising from transformations such as gate substitution and reordering, which preserve structural descriptors but alter execution behavior. However, because it depends on execution or simulation, it introduces computational overhead, particularly for circuits with large qubit counts \cite{pednault2018breaking}, reflecting the inherent complexity of evaluating quantum state behavior \cite{aaronson2016complexity}.

\subsection{Interaction Graph Semantic-Logical Score (IGS)}

The IGS, referred to as IGS-L in the implementation and code, is used consistently as IGS throughout this paper. To capture deviations beyond global structure, each circuit is represented as a labeled directed acyclic graph $G(C) = (V, E)$, where nodes correspond to quantum operations and edges represent dependencies induced by shared qubit usage and execution order \cite{itoko2020optimization,hagberg2008networkx}. This representation enables analysis of both structural dependencies and gate-level semantics without requiring circuit execution.

The interaction-level discrepancy is defined as:
\begin{equation}
\begin{aligned}
\Delta_{\text{IGS}}(C, C_{\text{ref}})
&= w_e D_{\text{edge}} + w_n D_{\text{node}} + w_o D_{\text{order}} \\
&\quad + w_i D_{\text{inter}} + w_u D_{\text{usage}}
\end{aligned}
\end{equation}

Each term $D_{\cdot} \in [0,1]$ represents a normalized interaction-level discrepancy, with weights satisfying $w_e + w_n + w_o + w_i + w_u = 1$. Weights are assigned based on sensitivity to interaction-level deviations, with higher emphasis on node features that encode gate type and parameters. To enhance semantic representation, unitary-based fingerprints are incorporated to distinguish operations with similar structure but different transformations \cite{nielsen2010quantum}. Interaction and ordering terms capture multi-qubit dependencies, while topology is assigned lower weight due to its limited sensitivity to structure-preserving transformations \cite{maslov2008quantum,amy2013meet}. A dedicated component captures anomalous qubit usage patterns, with weights initialized as $w_e = 0.15$, $w_n = 0.35$, $w_o = 0.20$, $w_i = 0.20$, and $w_u = 0.10$.

The Interaction Graph Semantic-Logical Score is defined as $\mathrm{IGS}(C, C_{\text{ref}}) = 1 - \Delta_{\text{IGS}}(C, C_{\text{ref}})$. The components capture complementary interaction-level properties, including dependency structure ($D_{\text{edge}}$) through graph topology, node semantics ($D_{\text{node}}$) via enriched features and unitary fingerprints, execution ordering ($D_{\text{order}}$), interaction patterns ($D_{\text{inter}}$) across multi-qubit relationships, and qubit usage ($D_{\text{usage}}$) to detect anomalous activity such as operations on idle qubits. These terms model differences in causal dependencies, operation characteristics, and qubit-level behavior. IGS operates in a pre-execution setting, providing an interaction-level perspective that complements structural and behavioral analysis of circuit integrity.

\subsection{Anomaly Model and Framework}

To evaluate the proposed metrics, controlled anomaly injection is used to generate perturbed circuit instances, simulating both unintended transformations (e.g., compilation artifacts) and adversarial modifications. The anomaly set includes gate deletion and insertion (structural changes), gate substitution and reordering (structure-preserving semantic changes), qubit swaps (interaction-level variation), and trojan operations that introduce hidden activity on idle qubits. These anomaly types introduce variations across structural, interaction-level, and behavioral dimensions, enabling systematic evaluation of metric sensitivity under diverse perturbation scenarios.

\begin{figure*}[!t]
    \centering
    \includegraphics[width=\textwidth]{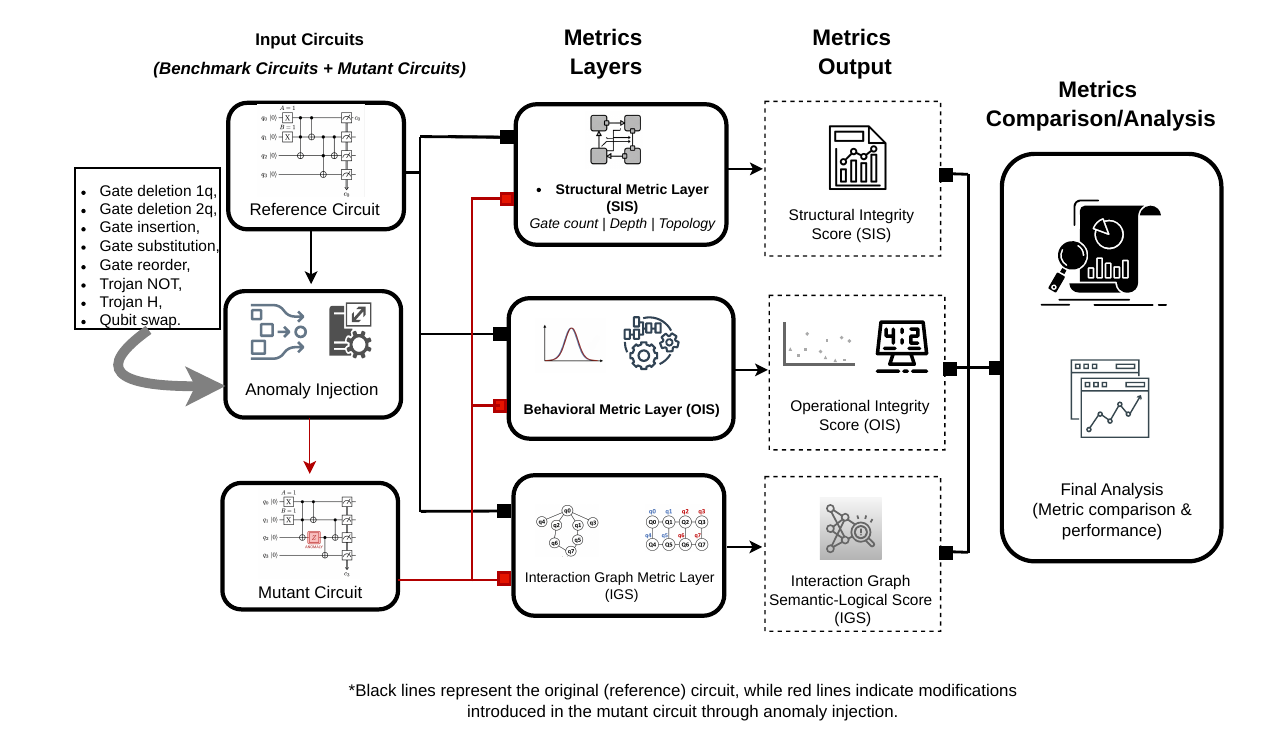}
    \caption{Multi-layer quantum circuit integrity framework using SIS, IGS, and OIS. Black lines show the reference circuit; red lines indicate anomalies.}
    \label{fig:framework}
\end{figure*}

Fig.~\ref{fig:framework} illustrates the overall workflow of the proposed multi-layer integrity validation framework. Given a reference circuit and its corresponding perturbed variants, generated through controlled anomaly injection, the framework evaluates circuit integrity across three complementary layers: structural, interaction-level, and behavioral.

The SIS captures global structural deviations using aggregated circuit descriptors such as gate count, depth, and topology. The IGS operates at the interaction level, modeling circuits as dependency graphs to capture ordering, semantic, and multi-qubit relationships. The OIS provides a behavioral perspective by measuring divergence between output distributions obtained through execution or simulation.

These three metrics are evaluated independently and produce separate outputs, preserving interpretability across different dimensions of analysis. SIS and IGS operate in a pre-execution setting, enabling efficient large-scale screening, while OIS provides execution-based validation of functional behavior.

The combination of these layers enables a broader assessment of circuit integrity. In particular, it allows identification of cases where structural similarity does not imply behavioral equivalence, as well as cases where interaction-level discrepancies explain differences between structural and behavioral observations. This multi-layer design provides a unified and interpretable framework for analyzing circuit deviations across structural, interaction-level, and behavioral domains.

From a security perspective, this design is motivated by scenarios in which circuits may be modified during compilation, optimization, or transmission. Such modifications may preserve structural properties while altering interaction patterns or execution behavior, making them difficult to detect using a single validation layer \cite{xu2024fault,roy2024trojan,das2024taxonomy}. The anomaly taxonomy reflects these potential threat patterns, and the three-layer framework enables detection across structural, interaction-level, and behavioral representations. This multi-layer perspective provides additional evidence for identifying integrity deviations in quantum circuit workflows, particularly in distributed or cloud-based environments \cite{xu2024fault,roy2024trojan}.



\begin{table}[t]
\caption{Theoretical grounding of metric components}
\label{tab:metric_foundations}
\centering
\begin{tabular}{p{2.5cm}p{3.6cm}p{1.6cm}}
\hline
\textbf{Component} & \textbf{Description} & \textbf{Reference} \\
\hline
Gate count & Circuit size and operation count & \cite{li2023qasmbench}  \\
Circuit depth & Sequential execution complexity & \cite{zulehner2018mapping} \\
Two-qubit gates & Entanglement and hardware cost & \cite{zulehner2018mapping} \\
Topology & Dependency structure (DAG) & \cite{itoko2020optimization} \\
Output distance & Behavioral difference via JSD & \cite{lin1991divergence,kullback1951information} \\
Graph representation & Interaction modeling & \cite{itoko2020optimization,hagberg2008networkx} \\
\hline
\end{tabular}
\end{table}

\section{Experimental Setup}

The experimental evaluation was conducted on QASM-based benchmark circuits organized into \textit{small}, \textit{medium}, and \textit{large} groups. Circuits were sourced from the QASMBench benchmark suite \cite{li2023qasmbench} and filtered to retain only instances with at most 40 qubits and 2000 gates. After filtering, the benchmark contained 133 circuits, including 76 small, 37 medium, and 20 large circuits. All experiments used a fixed random seed of 42 to ensure reproducibility.

Anomaly severity was evaluated at three levels, namely 0.1, 0.3, and 0.6. To ensure consistency across all evaluations, the same anomaly definitions and injection procedures were applied to all metrics. The SIS was computed using equal weights over four structural components: gate count, circuit depth, two-qubit gate count, and DAG-topology deviation~\cite{zulehner2018mapping}. The OIS was computed using the Qiskit Aer simulator backend with 1024 measurement shots \cite{abraham2019qiskit}, providing a behavioral reference based on output distributions. To control simulation cost, OIS evaluation was restricted to circuits with up to 14 qubits, consistent with exponential simulation constraints \cite{nielsen2010quantum}. 

To ensure separability between anomaly classes, the benchmark employed an orthogonal anomaly taxonomy consisting of eight perturbation types: \textit{single-qubit gate deletion}, \textit{two-qubit gate deletion}, \textit{gate insertion}, \textit{gate substitution}, \textit{gate reordering}, \textit{Trojan NOT}, \textit{Trojan H}, and \textit{qubit swap}. These perturbations are derived from mutation operators, fault injection strategies, and circuit transformation models studied in prior work on quantum software testing and hardware security \cite{mendiluze2021muskit, mendiluze2025empirical, xu2024fault, roy2024trojan, das2024taxonomy, maslov2008quantum, amy2013meet}. Each anomaly type was designed to primarily affect a distinct structural or interaction-level property, following established fault and transformation models in quantum circuits \cite{bera2015diagnosis,maslov2008quantum,amy2013meet}. Fixed-anomaly experiments applied a single minimal perturbation per circuit, while severity-based experiments scaled perturbation magnitude according to the specified severity factor.

All metrics were evaluated on a shared dataset derived from the QASMBench benchmark \cite{li2023qasmbench} using consistent filtering thresholds, anomaly definitions, severity levels, and random seed. The SIS and OIS benchmarks were implemented within a unified experimental pipeline, producing both fixed-anomaly and severity-based datasets. The IGS metric was evaluated using a companion pipeline with identical experimental settings to ensure comparability.

For IGS, each circuit was represented as a labeled interaction graph derived from its DAG structure \cite{itoko2020optimization}. Node representations combined gate-family encoding, a six-dimensional unitary-based feature vector, and qubit-level statistics. The six-dimensional unitary-based feature vector encodes gate-level characteristics derived from the corresponding unitary matrices, including parameter information (e.g., rotation angles), gate-type encoding (e.g., X, Y, Z, H), and aggregated summaries of real and imaginary components \cite{nielsen2010quantum}. These features provide a compact representation for interaction-level comparison without requiring circuit execution.

IGS was computed using five pre-execution components: topology difference, node semantic difference, gate-order difference, two-qubit interaction difference, and qubit-usage difference. To support scalability, the implementation incorporated reference-graph caching and a linear-time shared-qubit edge construction strategy, consistent with graph-processing practices \cite{hagberg2008networkx}. Qubit-usage monitoring was included to capture deviations associated with trojan-style perturbations \cite{roy2024trojan, das2024taxonomy}. OIS was used only as an external behavioral reference and was not incorporated into the computation of IGS.

Overall, the three metrics were evaluated under a unified experimental protocol, with each metric capturing a distinct aspect of circuit integrity: SIS reflects structural properties, OIS provides behavioral validation through execution, and IGS models interaction-level characteristics in a pre-execution setting. This design ensures that comparisons across metrics reflect differences in representation rather than differences in experimental conditions.
To support reproducibility, the complete implementation, datasets, and analysis scripts are publicly available in the replication package at \url{https://github.com/C2-Q/Q-SYNTRA}.

\subsection{Scope of Empirical Evaluation}

This study focuses on three implemented circuit-level integrity metrics: SIS, OIS, 
and IGS. These metrics are compared as complementary perspectives on circuit 
deviation rather than against separately implemented external baseline algorithms. 
SIS represents a global structural view, OIS represents an execution-based 
behavioral view, and IGS represents a pre-execution interaction-level view.

The comparison is designed to examine whether structural, interaction-level, and 
behavioral indicators respond differently to the same controlled anomalies. This is 
important because a circuit may remain structurally similar to its reference while 
still showing interaction-level or behavioral deviation. The evaluation uses the 
same benchmark circuits, anomaly definitions, severity levels, filtering criteria, 
and random seed across all three metrics to support a fair within-framework 
comparison.

Existing methods such as graph-edit comparison, global structural descriptors, state fidelity, and output-distribution divergence are relevant reference points for understanding the design space of circuit comparison approaches. Graph edit distance (GED) is a common graph-comparison measure that quantifies structural dissimilarity through edit operations between graph representations \cite{bunke1997graph,gao2010}. In this study, the empirical evaluation is scoped to the three implemented metrics: SIS, OIS, and IGS. Their complementary behavior under controlled anomaly injection is the central focus of the evaluation.

Table~\ref{tab:comparison_perspectives} summarizes the conceptual position of 
common circuit-comparison perspectives alongside the three evaluated metrics. 
Methods not marked in the Evaluated column are included to clarify the broader 
design space and are not separately implemented or empirically compared in this 
work.

\begin{table}[t]
\centering
\caption{Conceptual Positioning of Circuit Comparison Perspectives}
\label{tab:comparison_perspectives}
\renewcommand{\arraystretch}{1.2}
\scriptsize
\setlength{\tabcolsep}{2.5pt}
\begin{tabular}{p{1.75cm}ccccc}
\hline
\textbf{Method} & \textbf{Struct.} & \textbf{Behav.} & 
\textbf{Pre} & \textbf{Inter.} & \textbf{Eval.} \\
\hline
GED       
    & \checkmark & \xmark     & \checkmark & \xmark     & \xmark \\
Global features        
    & \checkmark & \xmark     & \checkmark & \xmark     & \xmark \\
State fidelity                  
    & \xmark     & \checkmark & \xmark     & \xmark     & \xmark \\
JSD 
    & \xmark     & \checkmark & \xmark     & \xmark     & \xmark \\
\hline
SIS                             
    & \checkmark & \xmark     & \checkmark & \xmark     & \checkmark \\
OIS ($1-\mathrm{JSD}$)          
    & \xmark     & \checkmark & \xmark     & \xmark     & \checkmark \\
IGS                             
    & \checkmark & \xmark     & \checkmark & \checkmark & \checkmark \\
\hline
\textbf{SIS+OIS+IGS}        
    & \checkmark & \checkmark & \textit{Partial} & \checkmark & \checkmark \\
\hline
\end{tabular}
\begin{flushleft}
\footnotesize{
\checkmark\ indicates that the perspective is supported; 
\xmark\ indicates that it is not. 
\textit{Partial} indicates that SIS and IGS operate before execution, while OIS requires simulation or execution. 
GED, global features, state fidelity, and standalone JSD are included as design-space reference points only and are not separately implemented or empirically compared in this study.
}
\end{flushleft}
\end{table}


\subsection{Implementation Details and Design Choices}

To keep the evaluation controlled and comparable across all experiments, a set of consistent implementation choices was applied. 


The anomaly design was intentionally kept structured. Each anomaly type was defined to affect a different aspect of the circuit, such as structure, ordering, or interaction patterns. This avoids overlap between anomaly effects and makes it easier to interpret which metric is responding to what. In the fixed-anomaly setting, only a minimal perturbation was applied to each circuit. In the severity-based setting, the same type of modification was scaled in a controlled way rather than introducing entirely different changes. This form of controlled perturbation is commonly used in anomaly detection and system evaluation to isolate the sensitivity of different detection mechanisms \cite{chandola2009anomalydetection}.

All metrics were evaluated against a single reference circuit for each instance. Anomalous versions were generated from this same baseline, so differences between SIS, OIS, and IGS reflect the behavior of the metrics themselves rather than differences in circuit selection.

Some factors were deliberately kept out of scope. No noise models or hardware-specific effects were included, and all simulations were performed using idealized backends. This follows standard practice in quantum circuit analysis, where functional behavior is first evaluated under noise-free conditions to isolate algorithmic effects before introducing hardware variability \cite{nielsen2010quantum}. The goal here was to observe how the metrics behave under clean conditions, where any deviation can be directly linked to the injected anomaly rather than randomness from noise.

Transpilation settings were also kept fixed, and no optimization-level changes were applied. This avoids the compiler modifying the circuit in unintended ways, such as reordering or simplifying gates, which could obscure the effect of injected anomalies. Prior work has shown that circuit transformations such as reordering and simplification can significantly alter structure and execution characteristics \cite{maslov2008quantum,amy2013meet}, making it important to control these effects during evaluation.

From an implementation perspective, reference simulation results for OIS were cached to avoid repeated execution of the same reference circuit. For IGS, the reference interaction graph was computed once per circuit and reused across anomaly cases. Graph construction was implemented using qubit-indexed structures, keeping the process close to linear in practice.

\section{Results}

This section evaluates how SIS, OIS, and IGS respond to controlled anomalies across quantum circuits of varying complexity and severity. We focus on three questions: how each metric responds to structural and semantic perturbations, whether OIS and IGS detect deviations missed by SIS, and how closely interaction-level signals align with behavioral outcomes.

\subsection{Structural Sensitivity of SIS}

\begin{figure*}[t]
    \centering
    \begin{subfigure}[t]{0.48\textwidth}
        \centering
        \includegraphics[width=\textwidth]{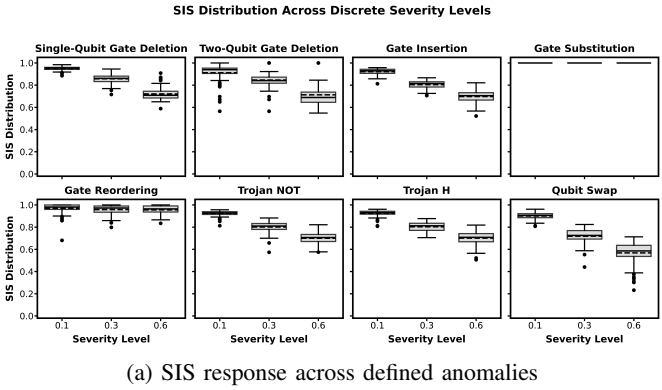}
        \caption{SIS response across defined anomalies}
        \label{fig:sis severity}
    \end{subfigure}
    \hfill
    \begin{subfigure}[t]{0.48\textwidth}
        \centering
        \includegraphics[width=\textwidth]{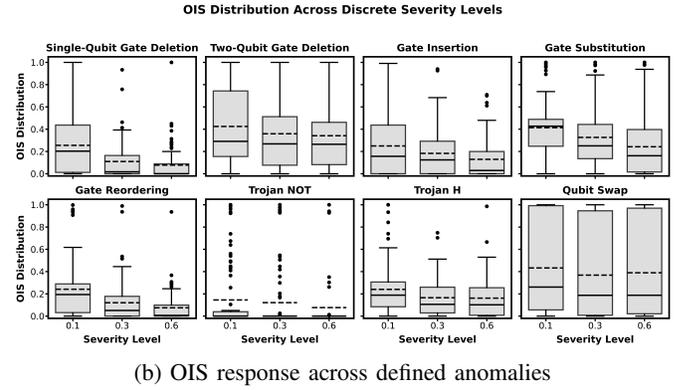}
        \caption{OIS response across defined anomalies}
        \label{fig:ois severity}
    \end{subfigure}

    \caption{Distribution of SIS and OIS across anomaly severity levels.}
    \label{fig:boxplots_combined}
\end{figure*}

SIS reliably detects anomalies that directly alter circuit structure, such as gate insertion, deletion, and depth modification. As shown in Fig.~\ref{fig:sis severity}, SIS decreases monotonically with increasing severity for gate deletion, gate insertion, trojan injection, and qubit swap anomalies, indicating that it reliably captures global structural degradation.

However, this sensitivity is not uniform across anomaly types. For gate substitution and gate reordering, SIS remains close to 1.0 across all severity levels, despite the presence of anomalous modifications. This reveals a clear structural blind-spot: transformations that preserve coarse global descriptors can evade detection even when the underlying computation has changed \cite{maslov2008quantum,amy2013meet}. In practical terms, this means that structural validation alone cannot be relied upon to establish circuit integrity.

\subsection{Behavioral Sensitivity of OIS}

The OIS captures behavioral deviations across all evaluated anomaly types. As illustrated in Fig.~\ref{fig:ois severity}, OIS consistently decreases with increasing severity, including in cases where SIS remains nearly unchanged. In particular, gate substitution and gate reordering produce substantial degradation in OIS despite minimal structural variation.

This directly shows that behavioral divergence can occur even when structural properties appear preserved \cite{bera2015diagnosis}. Compared with SIS, OIS therefore provides a more reliable indicator of functional deviation. However, the wide distributions observed for certain anomalies, particularly two-qubit gate deletion, Trojan-NOT, and qubit swap, indicate that behavioral impact is not uniform and depends on circuit structure and the location of the injected perturbation. The trade-off is cost: because OIS depends on execution or simulation, its use becomes less practical as circuit size grows \cite{nielsen2010quantum}. The result is a clear gap in the validation pipeline: SIS is efficient but incomplete, while OIS is informative but expensive.

\subsection{Interaction-Level Sensitivity of IGS}

The IGS evaluates circuit integrity by modeling dependencies and relationships between quantum operations in a graph-based representation. As shown in Fig.~\ref{fig:igs severity}, IGS exhibits a consistent decrease with increasing anomaly severity across all categories. This indicates that interaction-level structures degrade progressively as anomalies intensify, even in cases where global structural properties remain stable.

Compared with SIS, IGS is sensitive to changes in execution ordering, qubit interactions, and dependency structures. This allows it to capture deviations in scenarios such as gate reordering and trojan-based modifications, where structural metrics alone remain largely unchanged. At the same time, IGS does not replicate full behavioral analysis, and its responses remain distinct from those observed in OIS. The results position IGS as an intermediate metric that provides meaningful pre-execution insight into circuit deviations while maintaining computational efficiency.

\begin{figure}[ht]
    \centering
    \includegraphics[width=\columnwidth]{Fig_severity_IGS_boxplot_faceted.png}
    \caption{IGS response across defined anomalies.}
    \label{fig:igs severity}
\end{figure}

\subsection{Detection Capability in Structural Blind-Spots}

To explicitly evaluate cases where structural analysis is less sensitive, we analyze \textit{blind-spot scenarios} defined as instances where SIS remains high (SIS $\geq 0.95$) despite the presence of anomalies.
For the purpose of this analysis, OIS or IGS is considered to have detected an anomaly when the respective score falls below $0.95$, indicating a deviation of more than $5\%$ from perfect agreement with the reference circuit.

Across all experiments, a total of \textbf{569 blind-spot cases} were identified from \textbf{688 anomalous samples} in the severity-wise anomaly setting. These cases correspond to structurally indistinguishable circuits (SIS $\geq 0.95$), where structural validation fails to capture underlying deviations.

A severity-wise breakdown of blind-spot detection is presented below:

\begin{table}[h]
\centering
\caption{Detection Performance in Structural Blind-Spots (SIS $\geq 0.95$)}
\label{tab:blindspot_comparison}
\footnotesize
\setlength{\tabcolsep}{3.8pt}
\begin{tabular}{c c c c c}
\hline
\textbf{Severity} & \makecell{\textbf{Blind}\\\textbf{Cases}} & \makecell{\textbf{IGS}\\\textbf{Detected}} & \makecell{\textbf{OIS}\\\textbf{Detected}} & \makecell{\textbf{IGS / OIS}\\\textbf{(\%)}} \\
\hline
0.1 & 270 & 173 & 251 & 64.07 / 92.96 \\
0.3 & 151 & 113 & 142 & 74.83 / 94.04 \\
0.6 & 148 & 127 & 141 & 85.81 / 95.27 \\
\hline
\textbf{Total} & \textbf{569} & \textbf{413} & \textbf{534} & \textbf{72.58 / 93.85} \\
\hline
\end{tabular}
\end{table}

OIS successfully detected anomalies in \textbf{93.85\%} of blind-spot instances, while IGS detected anomalies in \textbf{72.58\%} of the same cases, reflecting their respective sensitivity to behavioral and interaction level deviations.

For both metrics, the detection rate increases consistently with severity. This indicates that deviations become more pronounced as anomaly intensity grows, even when structural properties remain unchanged. OIS captures the behavioral impact of these deviations through execution, whereas IGS identifies interaction level inconsistencies without requiring simulation.

These results highlight that structural validation captures global circuit properties effectively, but does not fully reflect interaction-level or behavioral deviations in blind-spot scenarios. In such cases, interaction-level and behavioral analyses provide complementary signals, extending the assessment beyond what can be inferred from structure alone.

\subsection{Cross-Metric Analysis and Efficiency}

\begin{figure*}[t]
\centering
\includegraphics[width=\textwidth]{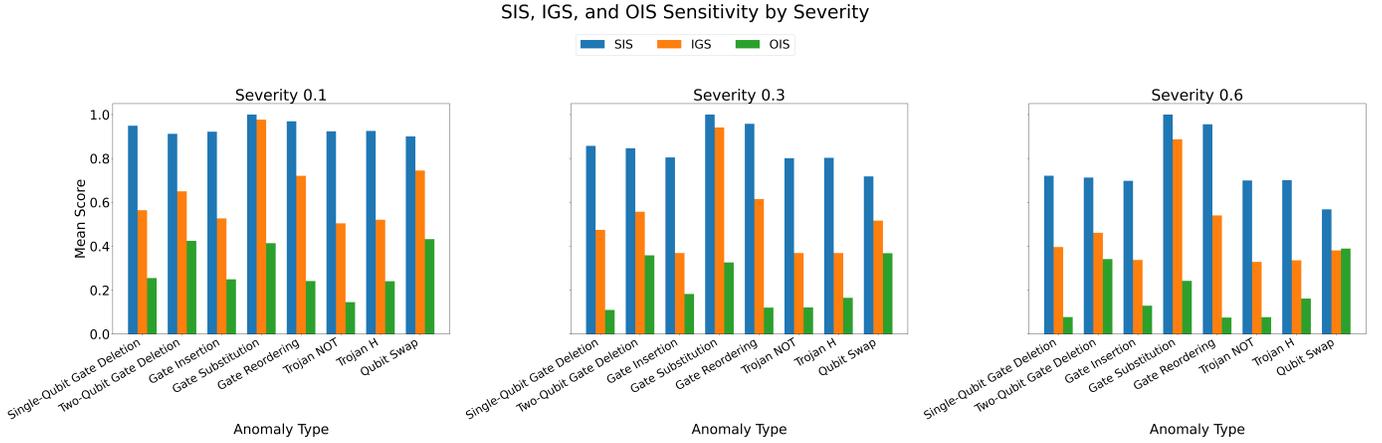}
\caption{SIS remains high for structure-preserving anomalies, while IGS and OIS degrade with increasing severity, capturing interaction level and behavioral deviations.}
\label{fig:sensitivity_3panel}
\end{figure*}

Fig.~\ref{fig:sensitivity_3panel} illustrates the behavior of SIS, OIS, and IGS across anomaly types for severity levels 0.1, 0.3, and 0.6.

SIS captures global circuit structure and responds clearly to direct structural changes while remaining relatively stable for structure-preserving transformations. In contrast, OIS shows a consistent decrease across all anomaly types as severity increases, reflecting sensitivity to changes in execution behavior. IGS follows a gradual, monotonic decline, capturing interaction-level differences that are not visible in SIS while remaining less variable than OIS. Each metric reflects a different aspect of circuit integrity, with distinct behavior across anomaly types and severity levels.

\textbf{Interaction–Behavior Relationship}

\begin{figure*}[t]
\centering
\includegraphics[width=\textwidth]{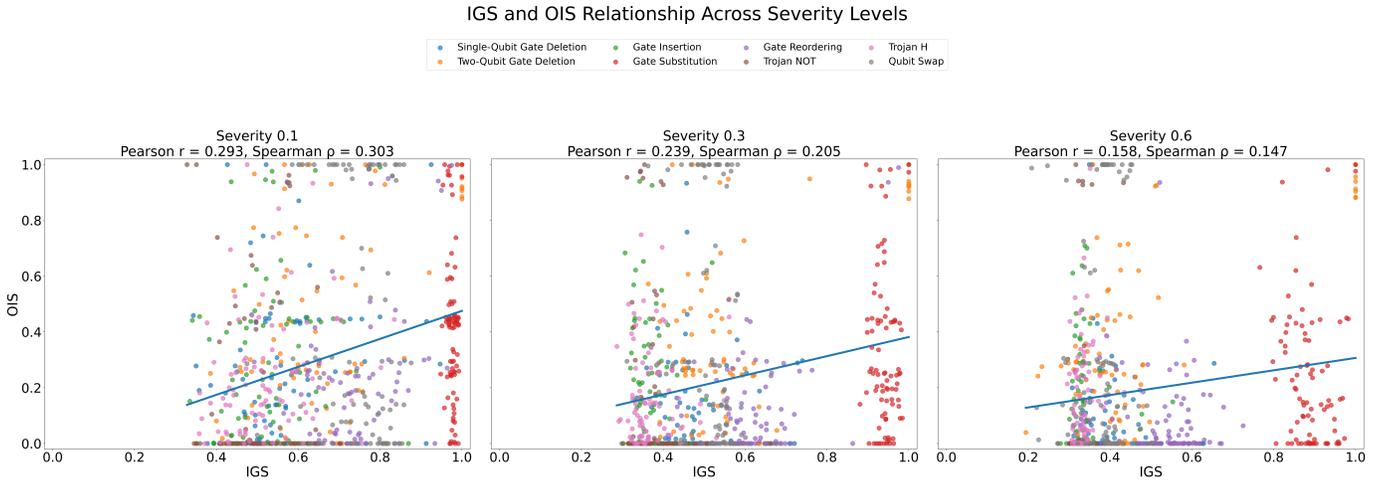}
\caption{IGS and OIS exhibit weak correlation across severity levels, confirming that interaction-level similarity does not reliably reflect behavioral equivalence.}
\label{fig:correlation_3panel}
\end{figure*}

The relationship between IGS and OIS is analyzed using correlation metrics across severity levels. Fig.~\ref{fig:correlation_3panel} shows the scatter distribution along with regression trends.

\begin{table}[h]
\centering
\caption{Correlation Between IGS and OIS Across Severity Levels}
\begin{tabular}{c|c|c|c|c}
\hline
\textbf{Severity} & \textbf{Pearson r} & \textbf{p-value} & \textbf{Spearman $\rho$} & \textbf{p-value} \\
\hline
0.1 & 0.293 & $4.00 \times 10^{-15}$ & 0.303 & $4.78 \times 10^{-16}$ \\
0.3 & 0.239 & $2.19 \times 10^{-10}$ & 0.205 & $6.21 \times 10^{-8}$ \\
0.6 & 0.158 & $3.18 \times 10^{-5}$  & 0.147 & $1.06 \times 10^{-4}$ \\
\hline
\end{tabular}
\label{tab:correlation}
\end{table}

Additionally, the overall correlation in the fixed anomaly setting is:

\begin{itemize}
    \item Pearson $r = 0.278$ ($p = 1.15 \times 10^{-13}$)
    \item Spearman $\rho = 0.267$ ($p = 1.11 \times 10^{-12}$)
\end{itemize}

The correlation between IGS and OIS remains weak across all settings and decreases as anomaly severity increases. Specifically, Pearson correlation drops from 0.293 at severity 0.1 to 0.158 at severity 0.6, with a similar trend observed for Spearman correlation.

This indicates that interaction-level similarity does not consistently align with behavioral similarity. As anomaly severity increases, the correlation between IGS and OIS decreases, indicating a progressively weaker association between the two metrics.

\textbf{Computational Efficiency}

\begin{figure*}[t]
\centering
\includegraphics[width=\textwidth]{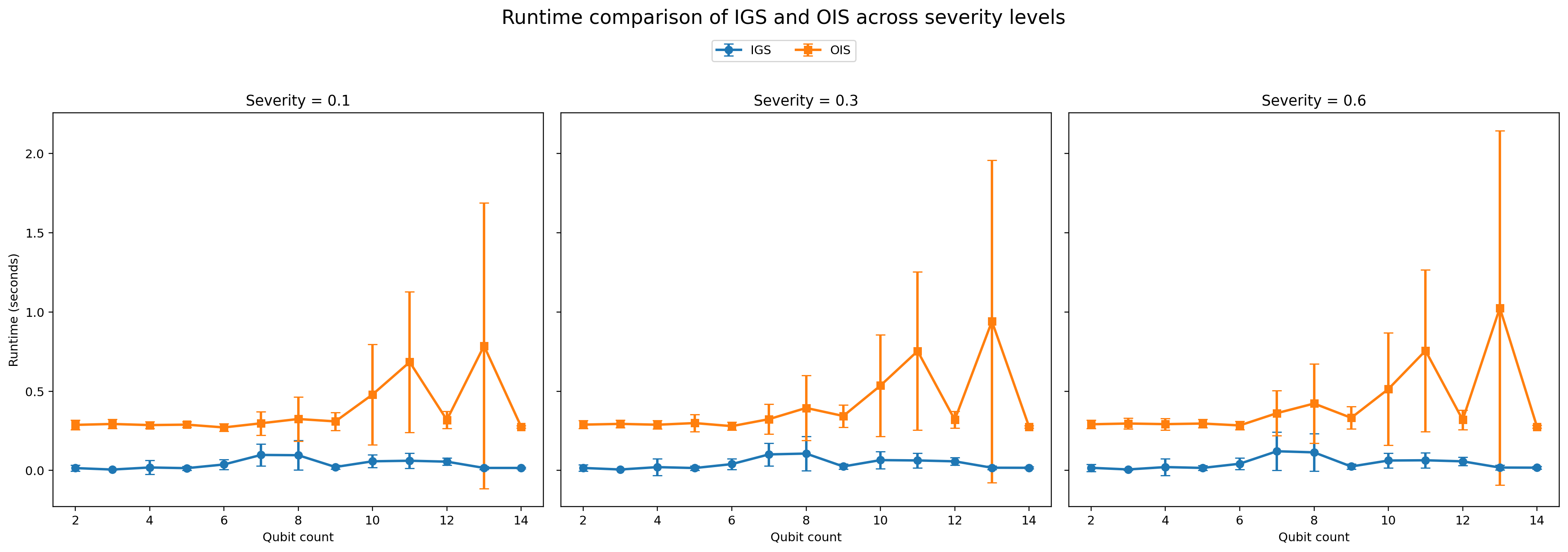}
\caption{IGS achieves stable and low runtime across qubit counts, while OIS incurs significantly higher and more variable computational cost.}
\label{fig:runtime_3panel}
\end{figure*}

Fig.~\ref{fig:runtime_3panel} compares the runtime of IGS and OIS across qubit counts for different severity levels.

IGS maintains consistently low and stable runtime across all configurations, with only a modest increase as the number of qubits grows and minimal variance across severity levels. In contrast, OIS exhibits substantially higher runtime and significantly greater variability, with both mean runtime and variance increasing sharply beyond 10 qubits. 

Notably, OIS runtime becomes increasingly unstable at higher qubit counts, with large error bars indicating sensitivity to circuit structure and execution conditions. This highlights the computational advantage of IGS as a low-runtime pre-execution metric in the evaluated setting, while reinforcing the cost limitations of behavior-based validation approaches.

\begin{tcolorbox}[colback=gray!5,colframe=black,title=Summary of Findings]
\small
SIS captures global structural properties, while OIS and IGS reflect behavioral and interaction-level deviations, respectively. Their responses vary across anomaly types and severity levels, highlighting distinct aspects of circuit integrity.

The three metrics provide complementary perspectives, enabling a more complete assessment of circuit integrity across structural, interaction-level, and behavioral dimensions.
\end{tcolorbox}

\section{Discussion}

The experimental results show a clear separation between structural, interaction-level, and behavioral perspectives of circuit integrity. As illustrated in Fig.~\ref{fig:sensitivity_3panel}, each metric responds differently to anomaly severity, indicating that deviations occur across multiple layers that cannot be captured by a single representation.

SIS captures global circuit properties such as topology and gate composition, but remains insensitive to structure-preserving transformations such as gate reordering and substitution (Fig.~\ref{fig:sis severity}). This highlights the limitation of relying on structural descriptors alone. In contrast, OIS captures deviations consistently across all anomaly types by reflecting changes in execution outcomes (Fig.~\ref{fig:ois severity}), although it requires simulation or hardware execution and does not localize the source of change.

IGS provides an intermediate view by modeling dependencies and operation ordering. It detects deviations not reflected in global structure (Fig.~\ref{fig:igs severity}), but does not fully align with behavioral changes. Correlation analysis (Fig.~\ref{fig:correlation_3panel}, Table~\ref{tab:correlation}) further shows that interaction-level similarity does not guarantee behavioral equivalence.

The blind-spot analysis highlights this divergence. In cases where circuits are structurally indistinguishable (SIS $\geq 0.95$), OIS reveals behavioral deviations, while IGS detects a substantial portion of these through interaction-level changes (Table~\ref{tab:blindspot_comparison}). This confirms that structural, interaction-level, and behavioral signals capture different aspects of circuit deviation.

From a computational perspective, interaction-level analysis remains relatively stable across circuit sizes, while behavioral evaluation incurs higher and more variable cost (Fig.~\ref{fig:runtime_3panel}), indicating a trade-off between efficiency and accuracy.

These observations suggest a practical validation workflow. SIS can be used for fast structural screening, IGS for pre-execution analysis of interaction-level inconsistencies, and OIS for selective confirmation of behavioral deviations. This staged approach leverages the strengths of each metric while managing computational cost.

These findings also have security implications. Modifications that preserve structural properties may still alter execution behavior, making them difficult to detect using structural validation alone. The observed mismatch between validation perspectives highlights the need for multi-level analysis in environments where circuits undergo external processing.

\section{Threats to Validation}

The evaluation is conducted in a controlled setting where anomalous circuits are generated from a single reference instance. This ensures consistent comparison between SIS, OIS, and IGS, but limits diversity in circuit structures and may not fully represent real-world workloads.

Experiments are performed under idealized conditions without noise models or hardware-specific effects. This isolates the impact of injected anomalies, but does not capture variability introduced by physical quantum devices, which may influence behavioral outcomes.

Transpilation settings are fixed and no optimization-level transformations are applied. While this prevents unintended compiler-induced changes, it does not reflect practical workflows where circuits undergo multiple compilation stages.
The OIS evaluation is restricted to circuits with up to 14 qubits due to the exponential cost of simulation. As a result, behavioral validation is limited to smaller circuits, and its applicability to larger-scale quantum workloads remains constrained. The IGS metric relies on manually defined component weights based on design considerations of sensitivity across structural and interaction-level features. These weights were not empirically optimized, and alternative configurations may influence sensitivity across anomaly types. In addition, the correlation between IGS and OIS is computed on aggregated results across circuits and severity levels, which can hide differences that are specific to certain anomaly types or circuit structures.

Finally, the anomaly taxonomy is limited to predefined perturbations. Although these are grounded in established models, they do not cover all possible real-world or adversarial modifications.

\section{Conclusion}

In this work, we investigated the limitations of single-perspective validation methods for quantum circuits by analyzing structural, interaction-level, and behavioral integrity metrics. We demonstrated that each metric captures a different aspect of circuit deviation, and that none of them alone provides a complete assessment of circuit correctness.

A key result of this study is the mismatch between these perspectives. In structural blind-spot cases (SIS $\geq 0.95$), where structural analysis fails to identify anomalies, OIS detects 93.85\% of anomalous instances, while IGS detects 72.58\%. This result demonstrates that structural similarity does not guarantee behavioral correctness, and that interaction-level analysis only partially bridges this gap.

The significance of this work lies in establishing the need for multi-dimensional validation strategies for quantum circuits. By combining structural, interaction-level, and behavioral perspectives, broader validation pipelines can be developed.

For future work, we aim to improve the alignment between interaction-level metrics and behavioral outcomes. The complementary behavior of SIS, OIS, and IGS suggests that they could be combined into a unified scoring function, such as a weighted integrity index, to provide a single interpretable signal for automated circuit monitoring. We also plan to explore how such a framework can be applied efficiently in practical quantum computing workflows.

\section*{Acknowledgement}

This work was supported by the Business Finland project (24304955) SeQuSoS (Secure Quantum Software Systems). The authors acknowledge the SeQuSoS research group for providing a collaborative research environment.

The authors acknowledge the use of the ChatGPT tool for language refinement, including grammar correction and sentence structuring, during the preparation of this manuscript.

\bibliographystyle{IEEEtran}
\bibliography{references}

\end{document}